# PHASE-OF-FIRING CODE

Anna Cattani[a], Gaute T. Einevoll[b,c], Stefano Panzeri[a*]

[a] Center for Neuroscience and Cognitive Systems @UniTn, Istituto Italiano di Tecnologia, 38068 Rovereto, Italy

[b] Department of Mathematical Sciences and Technology, Norwegian University of Life Sciences, 1432 Ås, Norway

[c] Department of Physics, University of Oslo, 0316 Oslo, Norway

[*] Corresponding author: stefano.panzeri@iit.it

## DEFINITION

The phase-of-firing code is a neural coding scheme whereby neurons encode information using the time at which they fire spikes within a cycle of the ongoing oscillatory pattern of network activity. This coding scheme may allow neurons to use their temporal pattern of spikes to encode information that is not encoded in their firing rate.

## DETAILED DESCRIPTION

In this entry, we define phase-of-firing coding, and we outline its properties. We first review early results that suggested the importance of temporal coding, and then we elucidate the insights concerning the central role of phase-of-firing coding in the studies of encoding of spatial variables in the hippocampus and of natural visual stimuli in primary visual and auditory cortices.

The phase-of-firing code is a temporal coding scheme that uses the timing of spikes with respect to network fluctuations to encode information about some variables of interest (such as the identity of an external sensory stimulus) that is not encoded by the firing rates of these neurons (Perkel and Bullock, 1968).

Figure 1 shows a schematic representation of both a phase-of-firing code and of a firing-rate code (i.e. a code that uses only the total number of spikes fired by the neuron to transmit information). In the case of firing-rate code (top panel), two hypothetical stimuli can be distinguished from one another based on the neural response because each stimulus evokes a different number of spikes. In the case of phase-of firing coding (bottom panel) the two hypothetical stimuli elicit neural responses which can be discriminated based on which phase (or part of the cycle) of network oscillations they are fired at, although these two hypothetical stimuli elicit similar spike counts and thus cannot be discriminated using a rate code. Thus, in a phase-of-firing code the time at which spikes are fired, measured with respect the cycle of network oscillations, carries unique information that cannot be retrieved when ignoring the timing of spikes.

PHASE-OF-FIRING CODING IN HOPFIELD'S ANALOGUE PATTERN RECOGNITION MODEL

The computational advantages of coding information by the phase-of-firing were highlighted in a classic model of analogue pattern recognition proposed by Hopfield (Hopfield, 1995). Hopfield considered the analogue pattern recognition problem of recognizing the identity of the stimulus or of a mixture of stimuli, whatever their overall intensity. This resembles many important operations that have to be routinely performed by sensory systems, for example the robust discrimination of the identity of an odor or of a mixture of odors independently of its intensity (Hopfield, 1995). Hopfield's model incorporates a set of "encoding" neurons each representing a specific stimulus. Each encoding neuron receives a specific depolarizing input whenever the stimulus it encodes is present (for example, a particular odorant). Crucially, all the encoding neurons also receive a subthreshold oscillatory current common to all of them. In this way it is easy to produce encoding neurons that, whenever the stimulus they encode is present, fire a spike in advance of the maximum of the intrinsic subthreshold oscillation (Figure 2A,B). This is a form of phase-of-firing coding of stimulus identity. Under the assumption that the stimulus-specific depolarizing current received by each encoding neuron increases with the stimulus intensity, any stimulus mixture can be easily identified by a downstream neuron with a short membrane time constant (thus implementing coincidence detection) that receives the firing of the encoding neurons with a set of transmission delays. These transmission delays match the relative timing of the firing in the encoding neurons for the considered stimulus mixture to ensure that all such inputs arrive simultaneously to some of the downstream decoding neuron (Fig 2C). A logarithmic transformation between stimulus strength and input current to the encoding neuron ensures that the times of firing in anticipation of the oscillation maximum of the encoding neurons are all rigidly shifted by the same amount if the overall strength of the stimulus mixture changes. Hopfield's model shows that encoding information by phase-of-firing makes it easy to read out this information in a scale invariant way with a plausible downstream neural system.

Hopfield noted that the scale-invariant analogue matching problem is difficult to solve with a neuron using a firing-rate code. Even a "grandmother" firing-rate neuron that is tuned only to a specific mixture of stimuli by a set of synaptic weights that boosts a specific stimulus mixture, will respond to sub-optimal stimulus mixtures provided that they are presented with a large enough intensity.

The phase-of-firing coding idea advanced by Hopfield coding was influential in introducing the notion that some problems that cannot easily be solved by firing-rate based schemes may be more easily solved by spike-timing based neural representations. Another influential aspect of Hopfield's phase-of-firing proposal was that it highlighted the fact that certain possible coding schemes cannot be detected when measuring spikes from single neurons only. However, whether or not the olfactory system encodes odor information in the way proposed by Hopfield still remains to be experimentally established.

PHASE-OF-FIRING ENCODING OF SPATIAL NAVIGATION VARIABLES IN THE HIPPOCAMPUS

The activity of hippocampal neurons has long been implicated in the encoding of information of variables important for spatial navigation. In particular, so called "place cells" in regions CA1 and CA3 of the rat hippocampal formation have been reported to encode the spatial location of the animal by elevating their firing when the animal is in a small spatial region called the place field of that neuron (O'Keefe, 1976).

In a seminal study, O'Keefe and Recce (O'Keefe and Recce, 1993) considered how these cells encoded information. They first noted that, in contrast to the firing rate of place cells, the hippocampal field potential theta pattern of rhythmical waves (7-12 Hz) is better correlated with a class of movements that change the rat's location in an environment. In the light of this observation, they considered the timing at which place cells fired in relation to the theta cycle. They found that firing consistently began at a particular phase as the rat entered the field but then shifted in a systematic way when traversing the field, moving progressively forward on each theta cycle. The phase-of-firing was highly correlated with spatial location and less well correlated with temporal aspects of behavior, such as the time after place-field entry. They suggested that by using the phase relationship as well as the firing rate, place cells might improve the accuracy of their place coding.

Later and more detailed studies have tried to clarify the exact nature of the information carried by the theta phase-of-firing of hippocampal cells and its relationship with the information carried by firing rates. Single-cell studies using the global statistical relationship between theta phase and spike firing could identify a partial dissociation of phase and firing rate in response to the animal's location and speed of movement (Huxter et al., 2003) but could not determine whether these signals encode complementary information about separate external features. Other studies have also shown that both theta phase and rate reflect the amplitude of the cell's input and might convey redundant spatial information (Mehta et al., 2002, Harris et al., 2002, Harris, 2005). However, a population study using decoding techniques (Huxter et al., 2008) demonstrated that the phase-of-firing of hippocampal place cells encodes features related to the animal's spatial navigation that are not encoded by spike rates. Their data suggest that the best position prediction is obtained from spiking in the descending theta-cycle phase, while place cell firing on the ascending slope encodes the direction rats are heading to (Huxter et al., 2008). These results imply that considering the theta phase at which hippocampal spikes are fired affords a richness of representation of information about position and heading that would not be possible with the firing-rate information only.

Note that Hafting and colleagues (Hafting et al., 2008) observed such phase-of-fire coding also in so called grid cells of the entorhinal cortex, one synapse upstream from the hippocampus. As such coding was not blocked by inactivation of the hippocampus, it raises the possibility that the phase-of-firing coding for hippocampal place cells is inherited from enthorinal cortex.

PHASE-OF-FIRING ENCODING OF NATURAL STIMULI IN PRIMARY SENSORY AREAS

The results discussed above suggest the importance of phase-of-firing coding in the hippocampal formation. However, until recently it has been unclear whether phase coding represents a fundamental currency for cortical information exchange also in primary sensory representation and whether it is a sufficiently robust coding mechanism to represent complex stimuli.

We investigated this question by recording spiking activity and local field potentials (LFPs) from the primary visual cortex of anaesthetized macaques while binocularly presenting a color movie (Montemurro et al., 2008). (LFPs are a massed measure of neural activity obtained by low pass filtering the extracellular potential recorded with intracranial microelectrodes and capturing fluctuations in the input and intracortical processing of the local cortical network, including the overall effect of population synaptic potentials and other types of slow activity, such as spike afterpotentials and voltage-dependent membrane oscillations; see (Buzsaki et al., 2012, Einevoll et al., 2013) for recent reviews). We found that the presentation of naturalistic color movies elicited reliable responses across trials both for the spikes and for the phase of LFPs in low frequency bands, particularly in the delta [1-4 Hz] and theta [4-8 Hz] band. However, we often found that two different movies scenes elicited very similar firing rates and thus could not be discriminated using a rate code. However, often the spikes elicited in responses to scenes with similar rate were fired at a different LFP phase in each scene, thereby suggesting that the phase of firing conveyed information that could not be extracted from spike rate. We investigated this point quantitatively by computing the amount of mutual information about which movie scene was being presented that can be extracted by firing rates and by phase of firing respectively (mutual information is a principled measure of association between variables that is often used in neuroscience to measure the association between sensory stimuli and neural responses; see (Quian Quiroga and Panzeri, 2009) for a review). The phase-of-firing was indeed found to convey information that was not accessible from the spike rates alone: the overall information about the movie increased by more than 50% when the LFP phase-of-firing was considered. The extra information available in the phase-of-firing was crucial for the disambiguation between stimuli eliciting high spike rates of similar magnitude. Thus, phase coding may allow primary cortical neurons to represent several effective stimuli in an easily decodable format.

We then investigated whether phase-of-firing coding of naturalistic information was present also in auditory cortex of awake macaques during the presentation of natural sounds input (Kayser et al., 2009). We found that also in this case, the phase-of-firing code carried information about the natural sound above and beyond that available in firing rates of auditory cortical neurons, thereby confirming the above results obtained in visual cortex. However, in auditory cortex we also found that an additional type of spike time coding scheme was at work: there were millisecond precise temporal patterns of spikes that too carried information not available in firing rates defined over longer time periods (Kayser et al., 2009). An important question was thus whether millisecond scale spike patterns carry the same or different information that the phase of slow network oscillations at which they are fired. Using again information theoretic measures, we found that information in spike rates and patterns at millisecond timescales was complementary to the information in the phase-of-firing, as sound periods not discernable from spike rates or spike patterns alone could instead be discriminated by also measuring the phase-of-firing. Together, these findings suggest that the timing of spikes measured on a millisecond scale, firing rates on a tens to hundred of millisecond scale, and timing of spikes measured with respect to network fluctuations operating a frequencies of a few hertz in sensory cortices constitute largely complementary information channels during naturalistic stimulation, and that these codes can be combined into highly informative codes containing "multiplexed" sensory information.

One way in which a neuron may use multiplexed codes that carry independent information at each time scale is illustrated in the simplified schematic of Figure 3. This figure shows a cartoon of a neuron that represents information needed to discriminate odd (1 and 3) from even (3 and 4) stimuli with the fine-time scale differences of their temporal firing patterns (very small inter-spike intervals are used for stimuli 1,3 and slightly longer inter-spike intervals are used for stimulus 2,4).

This example neuron represents on a much coarser time scale the information needed to discriminate stimuli 1,2 from stimuli 3,4 by using the timing of spikes with respect to the network oscillation (stimuli 1,2 are coded with spikes fired in the descending "green" part of the wave, whereas stimuli 3,4 are coded with spikes fired in the ascending "yellow" part of the wave). This example neuron thus carries all the information needed to discriminate the four stimuli only when putting together the complementary information from these two codes each operating at a different time scale.

As a caveat, it is worth mentioning that the fact that several different neural codes make a complementary contribution to the information encoded in a given brain area does not necessarily imply that all these codes make a complementary contribution to behavioral decisions based on the sensory variables represented by this neural activity. It is possible that for example, a downstream behavioral readout mechanism is unable to access the information is some of these codes. It is thus of paramount importance that future works tests the impact of different components of multiplexed codes on behavioral decisions. A complementary impact on behavior of millisecond spike patterns and firing rates has been recently reported in somatosensory cortex (Zuo et al, 2015). However, a complementary impact of phase of firing and firing rates on behavioral decisions has yet to be reported, to our knowledge.

Finally, we also used auditory cortex recordings to investigate which kinds of sensory representations are more robust to environmental noise in the sensory input (Kayser et al., 2009). We presented naturalistic sounds either in their original form or mixed with naturalistic environmental noise, which was varied from trial to trial. Although stimulus information decreased with increasing noise for all codes, the information in the phase-of-firing with respect to slow oscillations (hence a multiplexed code) increased relative to the information in spike counts when increasing noise. This suggests that encoding information in spike times using an internal temporal reference frame, such as slow network oscillations, may be central to forming stable sensory representations in unreliable environments.

ACKNOWLEDGEMENTS

We acknowledge the financial support of the VISUALISE and SICODE projects of the Future and Emerging Technologies (FET) Programme within the Seventh Framework Programme for Research of the European Commission (FP7-ICT-2011.9.11; FP7-600954 and FP7-284553), and of the European Community's Seventh Framework Programme FP7/2007-2013 (PITN-GA-2011-290011) and the Autonomous Province of Trento (Call "Grandi Progetti 2012", project "Characterizing and improving brain mechanisms of attention – ATTEND"). The funders had no role in decision to publish or preparation of the manuscript.

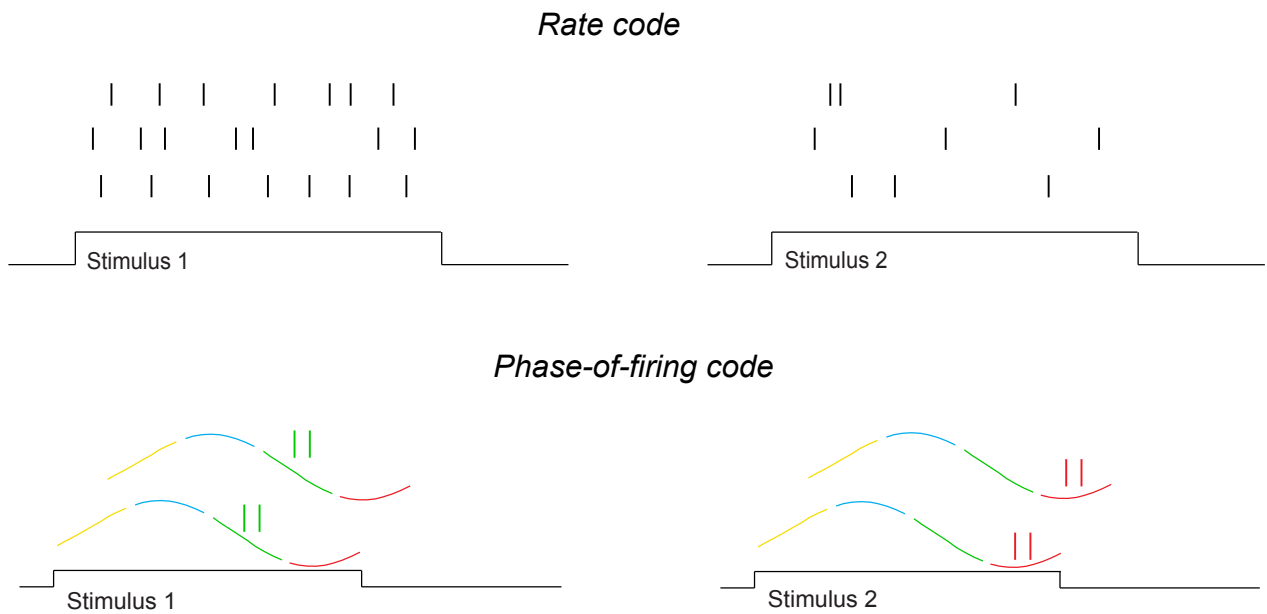

*Figure 1. Schematic cartoon of firing-rate code and phase-of-firing code.* In all panels, the x-axis represents peri-stimulus time, short vertical lines denote spike times, and each row corresponds to a different trial. Top panels show an example of firing-rate code where the two hypothetical stimuli (Left, Stimulus 1; Right, Stimulus 2) are encoded by the number of spikes fired by the neuron when the stimulus is presented. Bottom panels depict an example phase-of-firing code that carries information in terms of when spikes occur within a cycle of the ongoing oscillatory network activity. Here, the network oscillation phase is divided in four sectors depending on the phase and each of them is characterized by a different color. In this example, stimulus 1 (left) and stimulus 2 (right) cannot be distinguished by the firing rate, but by their respective phases of firing compared to the underlying oscillation (green versus red). Redrawn based on ideas presented in (Panzeri et al., 2010).

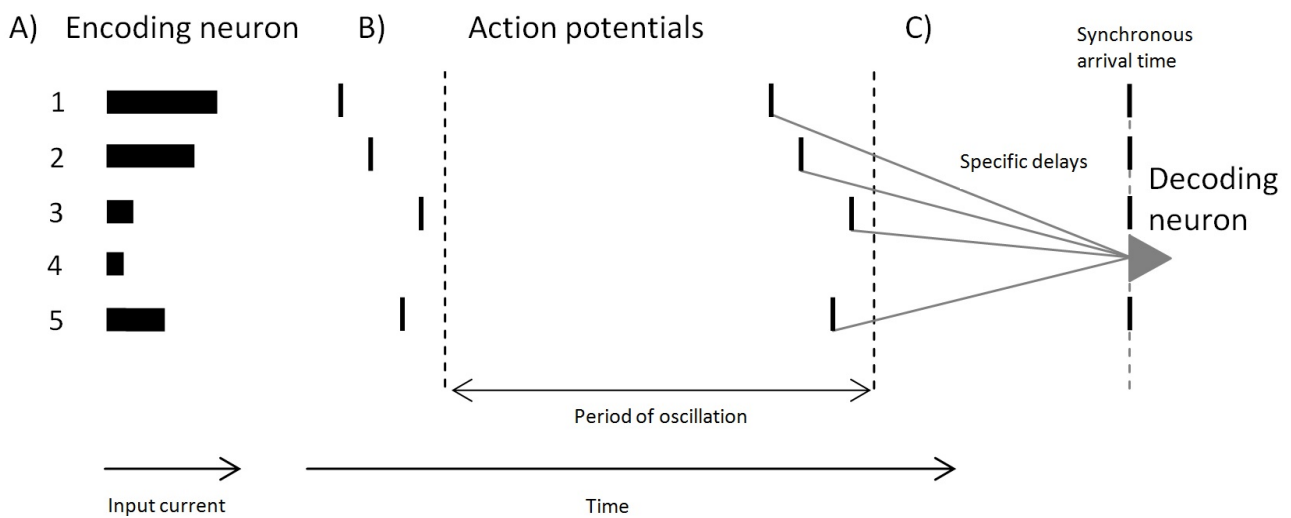

*Figure 2: Illustration of Hopfield phase-of-firing model of analogue mixture coding.* Panel B shows the action potentials of five different encoding neurons driven with different input current (corresponding to different analogues signal strengths represented by the length of the black bars in Panel A). Neuron 4 produces no action potential because its input current is too weak to drive that cell above threshold. The other neurons fire just once per oscillation cycle. The time each neuron fire in advance of the maximum of the subthreshold oscillation is larger for neurons driven with a greater input current. Panel C shows a scheme to read out the phase-of-firing coding of the stimulus mixture. The spikes generated by the encoding neurons reach out the soma of a decoding neuron at the same time because of suitably tuned differences in their conduction delays. The decoding neuron has a short membrane time constant and thus it fires only when it receives a sufficient number of simultaneous inputs. This only happens when the appropriate analogue stimulus mixture is presented. The logarithmic dependence on each stimulus's strength of the input current to each encoding neurons insure that an overall change of intensity of the analogue stimulus mixture does not change the relative time of arrival to the encoding neuron. Panels A and B adapted from (Hopfield, 1995).

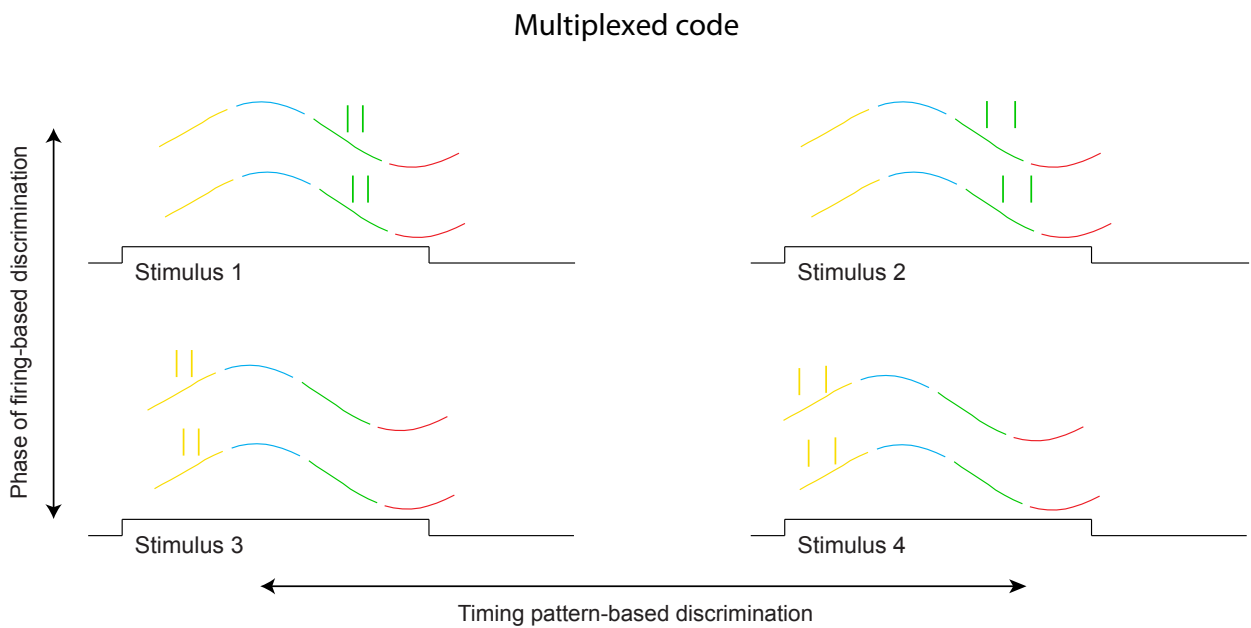

*Figure 3. Schematic cartoon of a multiplexed code*. In all panels, the x-axis represents peri-stimulus time, short vertical lines denote spike times and waves denote the concurrent network oscillation (measured e.g. as the Local Field Potential - LFP) recording together with the spikes, and each row corresponds to a different trial. Here, the network oscillation phase is divided in four sectors depending on the phase and each of them is characterized by a different color. In this example, stimulus 1 (left) and stimulus 2 (right) cannot be distinguished by the firing rate, but by their respective phases of firing compared to the underlying oscillation (green versus yellow). Redrawn from (Panzeri et al., 2010). Discrimination between odd (1 and 3) and even (2 and 4) stimuli is based on fine-precision temporal spike patterns (very small inter-spike intervals are used for stimuli 1,3 and slightly longer inter-spike intervals are used for stimulus 2,4). Discrimination between stimuli 1,2 from stimuli 3,4 is based on a phase-of-firing code based on oscillations at much slower time scales than that used the temporal spike patterns described above (stimuli 1,2 are coded with spikes fired in the descending "green" part of the wave, whereas stimuli 3,4 are coded with spikes fired in the ascending "yellow" part of the wave). In this cartoon, the example neuron thus uses multiplexing to code this stimulus set: the four stimuli can only be discriminated when putting together the complementary information from these two codes each operating at a different time scale.

# CROSS-REFERENCES

Local field potential

Firing rate

Phase-shift precession

Applications of Information Theory to Analysis of Neural Data

Summary of Information Theoretic Quantities

Extracellular Potentials, Forward Modeling of